\documentclass[ aps,%
floatfix,%
final,%
notitlepage,%
oneside,%
onecolumn,%
nobibnotes,%
nofootinbib,%
superscriptaddress,%
showpacs,%
]%
{revtex4}

\usepackage{epsfig}
\usepackage{amsfonts}
\usepackage{amsmath}
\usepackage{epsfig}
\usepackage{graphics}

\newcommand{\JP}{J/\Psi}

\newcommand{\beq}{\begin{eqnarray}}\newcommand{\eeq}{\end{eqnarray}}
\newcommand{\beqa}{\begin{eqnarray*}}\newcommand{\eeqa}{\end{eqnarray*}}

\begin{document}

\title{ The study of leading twist light cone wave functions of $\JP$ meson.}
\author{V.V. Braguta}
\email{braguta@mail.ru}
\affiliation{Institute for High Energy Physics, Protvino, Russia}

\begin{abstract}
This paper is devoted to the study of leading twist light cone wave functions of $\JP$ meson. 
The moments of these wave functions have been calculated within three approaches: potential models, 
nonrelativistic QCD and QCD sum rules. Using the results obtained within these approaches 
the models for the light cone wave functions of leading twist have been proposed. Similarly to the 
wave function of $\eta_c$ meson the leading twist light cone wave functions of $\JP$ meson 
have very interesting properties at scales $\mu> m_c$: improvement of the 
accuracy of the model, appearance of relativistic tail and violation of nonrelativistic QCD velocity scaling rules. 
The last two properties are the properties of true leading twist light cone wave functions 
of $\JP$ meson.
\end{abstract}

\pacs{
12.38.-t,  
12.38.Bx,  
13.66.Bc,  
13.25.Gv 
}

\maketitle

\newcommand{\ins}[1]{\underline{#1}}
\newcommand{\subs}[2]{\underline{#2}}
\vspace*{-1cm}
\section{Introduction}

Commonly exclusive charmonium production at high energies is studied within 
nonrelativistic QCD (NRQCD) \cite{Bodwin:1994jh}. In the framework of this approach charmonium is considered as 
a bound state of quark-antiquark pair moving with small relative velocity $v \ll 1$. 
Due to the presence of small parameter $v$ the amplitude of charmonium production can be built 
as an expansion in relative velocity $v$. 

Thus in the framework NRQCD the amplitude of any process is a series in relative velocity $v$. Usually,
in the most of applications of NRQCD, the consideration is restricted by the leading order approximation 
in relative velocity. However, this approximation has two problems which make it unreliable. The first problem is connected with 
rather large value of relative velocity for charmonium: $v^2 \sim 0.3,~v \sim 0.5$. For this value 
of $v^2$ one can expect large contribution from relativistic corrections in any process. So in any 
process resummation of relativistic corrections should be done 
or one should prove that resummation of all terms is not crucial. The second problem is connected with
QCD radiative corrections. The point is that due to the presence of large energy scale $Q$ there appears 
large logarithmic terms $(\alpha_s \log Q/m_c )^n, ~ Q \gg m_c $ which can be even more 
important than relativistic corrections at sufficiently large energy ( $Q \sim 10$ GeV). 
So these terms should also be resummed. In principle, it is possible to resum large logarithms
in the NRQCD factorization framework \cite{Petrelli:1999rh, Bodwin:2005hm}, however 
such resummation is done rarely. 

The illustration of all mentioned facts is the process of double charmonium production in $e^+ e^-$ annihilation at B-factories, 
where leading order NRQCD predictions \cite{Braaten:2002fi, Liu:2002wq, Liu:2004ga} are approximately by an order of magnitude less than experimental 
results \cite{Abe:2004ww,Aubert:2005tj}. The calculation of QCD radiative corrections \cite{Zhang:2005ch} diminished this
disagreement but did not remove it. Probably the agreement with the experiments can be 
achieved if, in addition to QCD radiative corrections, relativistic corrections will be resummed \cite{Bodwin:2006ke}.

In addition to NRQCD, hard exclusive processes can be studied in the framework of light cone expansion 
formalism \cite{Lepage:1980fj, Chernyak:1983ej} where both problems mentioned above can be solved. Within 
light cone expansion formalism the amplitude is built as an expansion 
over inverse powers of characteristic energy of the process. Usually this approach is successfully applied 
to excusive production of light mesons \cite{Chernyak:1983ej}. However recently the application of light cone 
expansion formalism to double charmonium production \cite{Bondar:2004sv, Braguta:2006nf, Braguta:2005kr, Ma:2004qf} allowed one
to achieve good agreement with the experiments.

In the framework of light cone formalism the amplitude of some meson production in any hard process  can be written as 
a convolution of the hard part of the process, which can be calculated using perturbative QCD,
and process independent light cone wave function (LCWF) of this meson that parameterizes nonperturbative effects.
From this one can conclude that charmonium LCWFs are key ingredients of any hard exclusive process with charmonium production. 
In paper \cite{Braguta:2006wr} the leading twist light cone wave function of $\eta_c$ meson was studied.
This paper is devoted to the study of leading twist LCWFs of $\JP$ meson.

The paper is organized as follows. In the next section all definitions needed in our calculation will be given. 
In Section III the moments of LCWFs will be calculated in the framework of Buchmuller-Tye and Cornell potential models. 
Section IV is devoted to the calculation of the moments within NRQCD. QCD sum rules will be applied 
to the calculation of the moments in Section V. Using the results obtained in Sections III-V 
the models of LCWFs will be built in Section VI. In the last section the results of this paper will be summarized.

\section{Definitions.}

There are two leading twist light cone wave functions (LCWF) of $\JP$ meson. The first one is 
twist 2 LCWF of longitudinally polarized $\JP$ meson $\phi_L ( \xi, \mu)$. The second one is 
twist 2 LCWF of transversely polarized $\JP$ meson $\phi_T ( \xi, \mu)$.
These LCWFs can be defined as follows \cite{Chernyak:1983ej}
\beq
{\langle 0| {\bar Q} (z) \gamma_{\alpha}  [z,-z] Q(-z) | \JP(\epsilon_{\lambda=0} ,p) \rangle}_{\mu} &=&
f_{L} p_{\alpha} \int^1_{-1} d \xi \,e^{i(pz) \xi}
\phi_L ( \xi, \mu), \nonumber \\
{\langle 0| {\bar Q} (z) \sigma_{\alpha \beta}  [z,-z] Q(-z) | \JP(\epsilon_{\lambda=\pm 1} ,p) \rangle}_{\mu} &=&
f_{T} (\mu) ( \epsilon_{\alpha} p_{\beta} - \epsilon_{\beta} p_{\alpha} )  \int^1_{-1} d \xi \,e^{i(pz) \xi}
\phi_T ( \xi, \mu),
\label{lwf}
\eeq
where the following designations are used: $x_1, x_2$ are the parts of momentum of the whole meson carried by quark and antiquark 
correspondingly, $\xi = x_1 - x_2$, $p$ is a momentum of $\JP$ meson, $\mu$ is an energy scale. The factor $[z,-z]$, that makes matrix element (\ref{lwf})
gauge invariant, is defined as 
\beq
[z, -z] = P \exp[i g \int_{-z}^z d x^{\mu} A_{\mu} (x) ].
\eeq
The LCWFs $\phi_{L, T} (\xi, \mu)$ are normalized as
\beq
\int_{-1}^1 d \xi~ \phi_{L, T} (\xi, \mu) =1. 
\label{norm}
\eeq
With this normalization condition the constants $f_{T, L}$ are defined as
\beq
{\langle 0| {\bar Q} (0) \gamma_{\alpha}  Q(0) | \JP(\epsilon_{\lambda=0} ,p) \rangle} &=&  f_{L} \epsilon_{\lambda=0}, \nonumber \\ 
{\langle 0| {\bar Q} (0) \sigma_{\alpha \beta} Q(0) | \JP(\epsilon_{\lambda=\pm 1} ,p) \rangle}_{\mu} &=&  f_{T} (\mu) 
( \epsilon_{\alpha} p_{\beta} - \epsilon_{\beta} p_{\alpha} ).
\label{def}
\eeq
It should be noted here that the local current ${\bar Q} (0) \gamma_{\alpha}  Q(0)$ is renormalization group invariant. 
The local current ${\bar Q} (0) \sigma_{\alpha \beta}  Q(0)$ is not invariant. For this 
reason the constant $f_T(\mu)$ depends on scale $\mu$ as
\beq
f_T (\mu) = \biggl ( \frac {\alpha_s( \mu)} {\alpha_s( \mu_0)} \biggr )^{\frac {4} {3 b_0}} f_T (\mu_0),
\eeq
but the constant $f_L$ does not depend on scale. 

LCWFs $\phi_{L, T} (x, \mu)$ can be expanded \cite{Chernyak:1983ej} in Gegenbauer polynomials $C_n^{3/2}( \xi)$ as follows
\begin{eqnarray}
\phi_{L, T} (\xi, \mu) = \frac 3 4 (1 - \xi^2) \biggl [  1 + \sum_{n=2,4..} a^{L,T}_n(\mu) C_n^{3/2} ( \xi ) \biggr ].
\label{conf_exp}
\end{eqnarray}
At leading logarithmic accuracy the coefficients $a_n^{L, T}$ are renormalized multiplicatively 
\begin{eqnarray}
a_n^{L, T}(\mu) = \biggl ( \frac {\alpha_s( \mu)} {\alpha_s( \mu_0)} \biggr )^{ \epsilon_n^{L, T} / {b_0}} a_n^{L, T}(\mu_0),
\label{ren}
\end{eqnarray}
where 
\beq
\epsilon_n^L &=& \frac 4 3 \biggl ( 1- \frac 2 {(n+1) (n+2)} + 4 \sum_{j=2}^{n+1} \frac 1 j \biggr ), \nonumber \\
\epsilon_n^T &=& \frac 4 3 \biggl (  4 \sum_{j=2}^{n+1} \frac 1 j \biggr ), ~~ b_0= 11 - \frac 2 3 n_{\rm fl}.
\label{an_dim}
\eeq
It should be noted here that conformal expansions (\ref{conf_exp}) are  solution of Bethe-Salpeter equation 
with one gluon exchange kernel \cite{Lepage:1980fj}.

From equations (\ref{conf_exp})-(\ref{an_dim}) it is not difficult to see that at infinitely large energy scale $\mu \to \infty$ 
LCWFs $\phi_{T, L} (\xi, \mu)$ tends to the asymptotic form $\phi_{as} (\xi) = 3/4 ( 1- \xi^2 )$. But at energy 
scales accessible at current experiments the LCWFs $\phi_{T, L}( \xi, \mu)$ are far from their asymptotic forms. The main 
goal of this paper is to calculate the LCWFs $\phi_{L, T}(\xi, \mu)$ of $\JP$ meson.
These LCWFs will be parameterized by their moments at some scale:
\beq
\langle \xi^n_{L, T}  \rangle_{\mu} = \int_{-1}^1 d \xi ~ \xi^n \phi_{L, T} (\xi, \mu).
\eeq
It is worth noting that since $\JP$ meson has negative 
charge parity the LCWFs $\phi_{L, T} ( \xi, \mu)$ are $\xi$-even. Thus all odd moments 
$\langle \xi_{L, T}^{2 k+1} \rangle$ equal zero and one needs to calculate only even moments.

Below  the following formulas will be used in our calculation
\beq
\langle 0 | \bar Q \gamma_{\nu}  (i z^{\sigma}  {\overset {\leftrightarrow} {D}}_{\sigma} )^n Q| \JP(\epsilon_{\lambda=0} ,p) \rangle &=& 
f_{L} p_{\nu} (zp)^n \langle \xi^n_{L} \rangle, \nonumber \\
\langle 0 | \bar Q \sigma_{\mu \nu}  (i z^{\sigma}  {\overset {\leftrightarrow} {D}}_{\sigma} )^n Q| \JP(\epsilon_{\lambda= \pm 1} ,p) \rangle &=& 
f_{T} (\epsilon_{\mu} p_{\nu} - \epsilon_{\nu} p_{\mu}  )  (zp)^n \langle \xi^n_{T} \rangle,
\label{xin}
\eeq
where
\beq
{\overset {\leftrightarrow} {D}}={\overset {\rightarrow} {D}}-{\overset {\leftarrow} {D}}, ~~~ {\overset {\rightarrow} {D}} = {\overset {\rightarrow} {\partial}}- 
i g B^a (\lambda^a /2).
\eeq
These formulas can be obtained if one expands both sides of equations (\ref{lwf}). 

\section{ The moments in the framework of potential models. }

In potential models charmonium mesons are considered as a quark-antiquark system
bounded by some static potential. These models allow one to understand many properties 
of chamonium mesons. For instance, the spectrum of charmonium family can be well 
reproduced in the framework of potential models \cite{Swanson:2006st}. Due to this 
success one can hope that potential models can be applied to the calculation of charmonium 
equal time wave functions. 

Having  equal time wave function of $\JP$ meson in momentum space $\psi( {\bf k} )$ one can apply 
Brodsky-Huang-Lepage (BHL) \cite{Brodsky:1981jv} procedure 
and get the LCWFs of leading twist $\phi_{L, T}(\xi, \mu)$ using the following rule:
\beq
\phi_{L, T} (\xi, \mu) \sim \int^{{\bf k}_{\perp}^2< \mu^2} d^2 k_{\perp} \psi_c(x,{\bf k}_{\perp}),
\label{p1}
\eeq
where $\psi_c(x,{\bf k}_{\perp})$ can be  obtained from $\psi( {\bf k} )$ after the 
substitution 
\beq
{\bf k}_{\perp} \to {\bf k}_{\perp}, \quad k_z \to ( x_1 - x_2) \frac {M_0} 2, \quad M_0^2 = \frac {M_c^2 + {\bf k}_{\perp}^2 } {x_1 x_2}.
\label{sub}
\eeq
Here $M_c$ is a quark mass in potential model. 
In this paper equal time wave function $\psi( {\bf k})$ will be calculated in the framework of the potential models with 
Buchmuller-Tye \cite{Buchmuller:1980su} and Cornell potentials \cite{Eichten:1978tg}.
The parameters of Buchmuller-Tye potential model were taken from paper \cite{Buchmuller:1980su}. For Cornell potential $V(r) = -k/r+ r/a^2$
the calculation was carried our with the following set of parameters: $k=0.61,~ a=2.38 \mbox{~GeV}^{-1},~ M_c=1.84$ GeV \cite{Eichten:2002qv}.

In paper \cite{Braguta:2006wr} the moments of leading twist LCWF of $\eta_c$ meson  
were calculated within potential models with these potentials. At leading order approximation 
in relative velocity there is no difference between equal time wave functions of $\eta_c$ and $\JP$ 
mesons. In what follows the moments obtained in paper \cite{Braguta:2006wr} for the leading twist LCWF of $\eta_c$ meson 
equal to the moments of LCWFs $\phi_{L, T} (\xi, \mu)$ of $\JP$ meson. Within this approximation 
there is no difference between $\phi_{L} (\xi, \mu)$ and $\phi_{T} (\xi, \mu)$.

It is worth noting that in paper \cite{Bodwin:2006dm} the relations between the light cone wave functions 
and equal time wave functions of charmonium mesons in the rest frame were derived. The procedure proposed in paper \cite{Bodwin:2006dm}
is similar to BHL with the difference: in formula (\ref {p1}) one must make the substitution 
$d^2 k_{\perp} \to d^2 k_{\perp} \sqrt {{\bf k}^2+m_c^2}/(4 m_c x_1 x_2)$. But this substitution was derived at leading 
order approximation in relative velocity of quark-antiquark motion inside the charmonium. At this approximation 
${\bf k}^2 \sim O(v^2),~ 4 x_1 x_2 \sim 1+ O(v^2)$ and the substitution amounts to $d^2 k_{\perp} \to d^2 k_{\perp} (1 + O( v^2 ))$. 
Thus at leading order approximation applied in \cite{Bodwin:2006dm} these two approaches coincide. 

The results of paper \cite{Braguta:2006wr} are presented in Table I (see this paper for details). 
In second and third columns the moments calculated in the framework of 
Buchmuller-Tye and Cornell models are presented. It should be noted that the moments from Table I
were calculated at scale $\mu \sim 1.5$ GeV. It is seen that there is  good agreement between these two 
models.

It should be noted here that the larger the power of the moment the larger contribution form 
the end point regions ($x \sim 0$ and $x \sim 1$) it gets. From formulas (\ref{sub}) one sees
that the motion of quark-antiquark pair in this region is relativistic and cannot be considered 
reliably in the framework of potential models. Thus it is not possible to calculate higher moments 
within the potential models. Due to this fact the calculation of the moments has been restricted by few first moments.

\section{ The moments in the framework of NRQCD.  }

In paper \cite{Braguta:2006wr}  the relations that allow one to connect the 
moments of leading twist LCWF of $\eta_c$ meson with NRQCD matrix elements were derived
\beq
\langle \xi^2 \rangle = \frac 1 {3 M_{c}^2} \frac {\langle 0 | \chi^+ (i {\bf {\overset {\leftrightarrow} {D}}})^2 \psi | \eta_c \rangle} 
{\langle 0 | \chi^+  \psi | \eta_c \rangle} = \frac {\langle  v^2 \rangle} {3}, \\ \nonumber
\langle \xi^4 \rangle = \frac 1 {5 M_c^4} \frac {\langle 0 | \chi^+ (i {\bf {\overset {\leftrightarrow} {D}}})^4 \psi | \eta_c \rangle} 
{\langle 0 | \chi^+  \psi | \eta_c \rangle} = \frac {\langle v^4 \rangle } 5,  \\ \nonumber 
\langle \xi^6 \rangle = \frac 1 {7 M_c^6} \frac {\langle 0 | \chi^+ (i {\bf {\overset {\leftrightarrow} {D}}})^6 \psi | \eta_c \rangle} 
{\langle 0 | \chi^+  \psi | \eta_c \rangle} = \frac {\langle v^6 \rangle } 7,
\eeq
where $\psi$ and $\chi^+$ are Pauli spinor fields that annihilate a quark and an antiquark respectively,
$M_c = M_{J/\Psi} /2$, The moments are defined at scale $\mu \sim M_c$. 

These relations were derived at leading order approximation in relative velocity. However, 
as it was noted above at this approximation there is no difference between $\eta_c$ and $\JP$ 
mesons. Moreover, there is no difference between LCWFs $\phi_{L} (\xi, \mu)$ and $\phi_{T} (\xi, \mu)$.
So the values for the moments of LCWFs $\phi_{L} (\xi, \mu), \phi_{T} (\xi, \mu)$ can be taken
from paper \cite{Braguta:2006wr}. 

The results of the calculation of the moments  are presented in the fourth column of Table I. 
The central values of the moments and the errors due to the model uncertainty 
have been calculated according to the approach proposed in paper \cite{Bodwin:2006dn}. In addition to the
error shown in Table I there is an uncertainty due to higher order $v$ corrections. For the second moment 
one can expect that this error is about $30 \%$. For higher moments this error is larger.

It is seen from Table I that NRQCD predictions for the second and the fourth moments are in good agreement 
with potential model and there is disagreement for the moment $\langle \xi^6 \rangle$ between these two approaches. 
The cause of this disagreement is the fact noted above: due to the large contribution of 
relativistic motion of quark-antiquark pair inside quarkonium it is not possible to apply both 
approaches for higher moments. So one can expect that both approaches can be used for the estimation 
of the values of the second and the fourth moments. The predictions for the sixth 
and higher moments  become unreliable.

\begin{table}
$$\begin{array}{|c|c|c|c|c|c|}
\hline \langle \xi^n \rangle & \mbox{ Buchmuller-Tye } & \mbox{ Cornell } & \mbox{  NRQCD }
& \mbox {QCD sum rules} & \mbox {QCD sum rules}  \\
  & \mbox{ model  \cite{Buchmuller:1980su} } & \mbox{ model \cite{Eichten:1978tg}} & \mbox{\cite{Bodwin:2006dn} } & \phi_{L} (\xi, \mu) 
  & \phi_{T} (\xi, \mu) \\
\hline
n=2  & 0.086  & 0.084  &  0.075 \pm 0.011   & 0.070 \pm 0.007 & 0.072 \pm 0.007  \\
\hline
n=4  & 0.020 & 0.019 &   0.010 \pm 0.003  & 0.012 \pm 0.002 & 0.012 \pm 0.002  \\
\hline
n=6  & 0.0066 & 0.0066 &  0.0017 \pm 0.0007 & 0.0031 \pm 0.0008 & 0.0033 \pm 0.0007  \\
\hline
\end{array}$$
\caption{The moments of LCWFs $\phi_{L} (\xi, \mu), \phi_{T} (\xi, \mu)$ obtained within different approaches. 
In the second and third columns the moments calculated in the framework of 
Buchmuller-Tye and Cornell potential models are presented. In the fourth column NRQCD predictions for the moments are presented. 
In last two columns the results obtained within QCD sum rules are shown. }
\end{table}

\section{ The moments in the framework of QCD sum rules. }

In this section QCD sum rules \cite{Shifman:1978bx, Shifman:1978by}  will be applied to the calculation of the moments of LCWFs 
$\phi_{L} (\xi, \mu)$ and $\phi_{T} (\xi, \mu)$ \cite{Chernyak:1983ej, chernyak}. First let us consider 
LCWF $\phi_{L} (\xi, \mu)$. To calculate the moments of this LCWF one should consider two-point 
correlator:
\beq
\Pi_L (z,q) = i \int d^4 x e^{i q x} \langle 0| T J_0(x) J_n (0) |0 \rangle = (zq)^{n+2} \Pi_L (q^2), 
\label{corL} \\ \nonumber
J_0 (x) = \bar Q(x) \hat z  Q(x), ~~~ J_n(0) = \bar Q(0) \hat z (i z^{\rho} {\overset {\leftrightarrow} {D}_{\rho}} )^n  Q(0), ~~ z^2=0.
\eeq
It is not difficult to obtain sum rules for this correlator (for details see paper \cite{Braguta:2006wr}). 
\beq
\frac {f_{L}^2 \langle \xi^n_L \rangle}  { (M_{J/\Psi}^2+Q^2)^{m+1} } = 
\frac 1 {\pi} \int_{4 m_c^2}^{\infty} ds ~ \frac {\mbox{Im} \Pi_{\rm pert}(s)} {(s+Q^2)^{m+1} }  + \Pi^{(m)}_{\rm npert}(Q^2),
\label{smL}
\eeq
where perturbative and nonperturbative contributions to sum rules $\mbox {Im} \Pi_{\rm pert} (s)$, $\Pi^{(m)}_{\rm npert}(Q^2)$  
can be written as 
\beq
\mbox {Im} \Pi_{\rm pert} (s) = \frac {3} {8 \pi} v^{n+1}  (\frac 1 {n+1} - \frac {v^2} {n+3} ), ~~~~ v^2 = 1 - \frac {4 m_c^2} {s},
\label{pert}
\eeq
\beq
\label{power}
\Pi^{(m)}_{\rm npert} (Q^2) &=& \Pi_1^{(m)} (Q^2) + \Pi_2^{(m)} (Q^2)+\Pi_3^{(m)} (Q^2), \\ \nonumber
\Pi_1^{(m)} (Q^2) &=& \frac {\langle \alpha_s G^2 \rangle} {24 \pi }
(m+1) \int_{-1}^1 d \xi~  \biggl (\xi^n + \frac {n(n-1)} 4 \xi^{n-2} (1- \xi^2) \biggr )
\frac {(1- \xi^2)^{m+2}} {\bigl ( 4 m_c^2  + Q^2 (1- \xi^2 ) \bigr )^{m+2} }, \\ \nonumber
\Pi_2^{(m)} (Q^2) &=& - \frac {\langle \alpha_s G^2 \rangle} {6 \pi} m_c^2 ( m^2+ 3m +2)
\int_{-1}^1 d \xi~ 
  \xi^n \bigl ( 1+ 3 \xi^2 \bigr ) \frac {(1- \xi^2)^{m+1}} {\bigl ( 4 m_c^2  + Q^2 (1- \xi^2 ) \bigr )^{m+3} }, \\ \nonumber
\Pi_3^{(m)} (Q^2) &=&  \frac {\langle \alpha_s G^2 \rangle} {384 \pi }
(n^2-n) (m+1) \int_{-1}^1 d \xi~  \xi^{n-2} 
\frac {(1- \xi^2)^{m+3}} {\bigl ( 4 m_c^2  + Q^2 (1- \xi^2 ) \bigr )^{m+2} }.
\eeq
Here $Q^2=-q^2$, $m_c$ and $\langle {\alpha_s} G^2 \rangle$ are parameters of QCD sum rules. 

To calculate the moments of LCWF $\phi_{T} (\xi, \mu)$ one should consider the correlator:
\beq
\Pi_T (z,q) &=& i \int d^4 x e^{i q x} \langle 0| T J_{\mu}(x) J_n^{\mu} (0) |0 \rangle = (zq)^{n+2} \Pi_T (q^2), 
\label{corT} \\ \nonumber
J_{\mu} (x) &=& \bar Q(x) (\sigma_{\mu \nu}  z^{\nu})  Q(x), ~~~ J_n^{\mu}(0) = \bar Q(0) (\sigma^{\mu \nu}  z_{\nu}) (i z^{\rho} {\overset {\leftrightarrow} {D}_{\rho}} )^n  Q(0), ~~ z^2=0.
\eeq
The sum rules for this correlator can be written as 
\beq
\frac {f_{T}^2 \langle \xi^n_T \rangle}  { (m_{J/\Psi}^2+Q^2)^{m+1} } = 
\frac 1 {\pi} \int_{4 m_c^2}^{\infty} ds ~ \frac {\mbox{Im} \Pi_{\rm pert}(s)} {(s+Q^2)^{m+1} }  + \Pi^{(m)}_{\rm npert}(Q^2),
\label{smT}
\eeq
where perturbative and nonperturbative contributions to sum rules (\ref{smT}) are given by expressions (\ref{pert}), (\ref{power})
except that the expression for $\Pi_1^{(m)} (Q^2)$ should be replaced by 
\beq
\Pi_1^{(m)} (Q^2) &=& \frac {\langle \alpha_s G^2 \rangle} {24 \pi }
(m+1) \int_{-1}^1 d \xi~  \biggl (- \xi^n + \frac {n(n-1)} 4 \xi^{n-2} (1- \xi^2) \biggr )
\frac {(1- \xi^2)^{m+2}} {\bigl ( 4 m_c^2  + Q^2 (1- \xi^2 ) \bigr )^{m+2} }.
\eeq

In the original paper \cite{Shifman:1978by} 
the method QCD sum rules was applied at $Q^2=0$. However, as it was shown in paper \cite{Reinders:1984sr},
there is large contribution of higher dimensional operators at $Q^2=0$ which grows rapidly with $m$. 
To suppress this contribution  sum rules (\ref{smL}), (\ref{smT}) will be applied at $Q^2=4 m_c^2$. 

In the numerical analysis of QCD sum rules the values of parameters $m_c$ and $\langle \alpha_s G^2/ \pi \rangle$ will be 
taken from paper \cite{Reinders:1984sr}:
\beq
m_c = 1.24 \pm 0.02 ~\mbox {GeV}, ~~ \langle \frac {\alpha_s} {\pi} G^2 \rangle = 0.012 \pm 30 \% ~\mbox {GeV}^4.
\label{param}
\eeq
First sum rules (\ref{smL}), (\ref{smT}) will be applied to the calculation of the constants $f_{L, T}^2$. 
It is not difficult to express the constants $f_{L, T}^2$  from equations (\ref{smL}), (\ref{smT}) at $n=0$ as functions
of $m$. For too small values of $m$ ($m<m_1$) there are large contributions from  higher resonances 
and continuum which spoil sum rules (\ref{smL}), (\ref{smT}). Although for $m \gg m_1$ these contributions
are strongly suppressed, it is not possible to apply sum rules for too large $m$ ($m>m_2$) 
since the contribution arising from higher dimensional vacuum condensates rapidly grows with $m$ and invalidates 
approximation of this paper. If $m_1<m_2$ there is some region of applicability of sum rules (\ref{smL}), (\ref{smT}) 
$[m_1, m_2]$ where both resonance and  higher dimensional vacuum condensates contributions are not too large. 
Within this region 
$f_{L, T}^2$ as a functions of $m$ vary slowly and one can determine the values of these constants.
Applying approach described above one gets 
\beq
f_{L}^2 = 0.170 \pm 0.002 \pm 0.004 \pm 0.016 ~ \mbox {GeV}^2, \nonumber \\
f_{T}^2 = 0.167 \pm 0.002 \pm 0.003 \pm 0.016 ~ \mbox {GeV}^2.
\label{fpsi}
\eeq
The first error in (\ref{fpsi}) corresponds the variation of the constants $f_{L, T}^2$
within the region of stability. The second and the third errors in (\ref{fpsi}) correspond to the 
variation  of the gluon condensate $\langle \alpha_s G^2 \rangle$ and the mass $m_c$ within ranges (\ref{param}). From the 
results (\ref{fpsi}) one sees that the main errors in determination of the constants $f_{L, T}^2$ 
result from the variation of the parameter $m_c$. This fact represents well known property: 
high sensitivity of QCD sum rules to the mass parameter $m_c$.

Next let us consider the second moments $\langle \xi^2_{L, T} \rangle$ in the framework of QCD sum rules. One way 
to find the values of $\langle \xi^2_{L, T} \rangle$ is to determine the values of 
$f_{L, T}^2 \langle \xi_{L, T}^2 \rangle$ 
from sum rules (\ref{smL}), (\ref{smT}) at $n=2$ and then extract $\langle \xi^2_{L, T} \rangle$. 
However, as it was noted above, this approach suffers from high sensitivity of right side of equations (\ref{smL}), (\ref{smT}) to the variation 
of the parameter $m_c$. Moreover, the quantities $f_{L, T}^2 \langle \xi_{L, T}^2 \rangle$ include 
not only the errors in determination of  $\langle \xi^2_{L, T} \rangle$, 
but also the errors in $f_{L, T}^2,$. To remove these disadvantages   the ratios of sum rules at $n=2$ and $n=0$: 
$f_{L, T}^2 \langle \xi_{L, T}^2 \rangle / f_{L, T}^2$ will be considered.
The moments $\langle \xi^4_{L, T} \rangle, \langle \xi^6_{L, T} \rangle$ will be considered analogously. Applying
standard procedure one gets the moments of LCWF $\phi_{L} (\xi, \mu)$:
\beq
\label{resL}
\langle \xi_L^2 \rangle &=& ~0.070 \pm 0.002 \pm 0.007 \pm 0.002,  \\
\langle \xi_L^4 \rangle &=& ~0.012 \pm 0.001 \pm 0.002 \pm 0.001, \nonumber \\
\langle \xi_L^6 \rangle &=& ~0.0031 \pm 0.0002 \pm 0.0008 \pm 0.0002, \nonumber  
\eeq
and moments of LCWF $\phi_{T} (\xi, \mu)$
\beq
\label{resT}
\langle \xi_T^2 \rangle &=& ~0.072 \pm 0.002 \pm 0.007 \pm 0.002,  \\
\langle \xi_T^4 \rangle &=& ~0.012 \pm 0.001 \pm 0.002 \pm 0.001, \nonumber \\
\langle \xi_T^6 \rangle &=& ~0.0033 \pm 0.0002 \pm 0.0007 \pm 0.0003. \nonumber  
\eeq
The first error in (\ref{resL}), (\ref{resT}) corresponds the variation 
within the region of stability. The second and the third errors in (\ref{resL}), (\ref{resT}) correspond to the 
variation  of the gluon condensate $\langle \alpha_s G^2 \rangle$ and the mass $m_c$ within ranges (\ref{param}). It is 
seen that, as one expected, the sensitivity of the ratios $f_{L,T}^2 \langle \xi_{L,T}^n \rangle / f_{L,T}^2$ to the variation of 
$m_c$ is rather low. The main source of uncertainty is the variation of gluon condensate $\langle \alpha_s G^2 \rangle$.
In the fifth and sixth columns of Table I results (\ref{resL}), (\ref{resT}) are presented. The errors in Table I correspond 
to the main source of uncertainty --- the variation of gluon condensate $\langle \alpha_s G^2 \rangle$. 

In the calculations of the correlators (\ref{corL}), (\ref{corT}) characteristic virtuality of quark 
is $\sim (4 m_c^2 + Q^2)/m \sim m_c^2$. So the values of the moments (\ref{resL}), (\ref{resT})  are defined at 
scale $\sim m_c^2$. 

From Table I it is seen that the larger the number of the moment $n$ the larger the 
uncertainty due to the variation of vacuum gluon condensate. This property is a consequence 
of the fact that the role of power corrections in the sum rules (\ref{smL}), (\ref{smT}) grows with $n$.
From this one can conclude that there is considerable nonperturbative contribution to 
the moments $\langle \xi_{L,T}^n \rangle $ with large $n$ what means that nonperturbative effects are very 
important in relativistic motion of quark-antiquark pair inside the meson. The second 
important contribution to QCD sum rules (\ref{smL}), (\ref{smT}) at large $n$ is QCD radiative corrections
to perturbative part $\Pi_{\rm pert}(Q^2)$.
Unfortunately today one does not know the expression for these corrections and for 
this reason they are not included to sum sules (\ref{smL}), (\ref{smT}). One can only say 
that these corrections grow with $n$ and, probably, the size of radiative corrections to 
the ratios $f_{L, T}^2 \langle \xi_{L, T}^n \rangle / f_{L, T}^2$ is not too big for not too large $n$. 
Thus one can expect that QCD radiative corrections will not change dramatically the results 
for the moments  $n=2$ and $n=4$. But the radiative corrections to  $\langle \xi_{L, T}^6 \rangle$ may be important. 

It is interesting to compare the moments of leading twist LCWF of $\eta_c$ meson calculated 
in paper \cite{Braguta:2006wr}
\beq
\label{reseta}
\langle \xi_{\eta_c}^2 \rangle &=& ~0.070 \pm 0.002 \pm 0.007 \pm 0.003,  \\
\langle \xi_{\eta_c}^4 \rangle &=& ~0.012 \pm 0.001 \pm 0.002 \pm 0.001, \nonumber \\
\langle \xi_{\eta_c}^6 \rangle &=& ~0.0032 \pm 0.0002 \pm 0.0009 \pm 0.0003, \nonumber  
\eeq
with the results of this section. It is seen that there is no significant difference 
between the moments of leading twist LCWF of $\eta_c$ meson and the moments of LCWFs 
$\phi_{L} (\xi, \mu), \phi_{T} (\xi, \mu)$. The difference is within the error of the 
calculation. 

\newpage
\begin{figure}[ph]
\begin{picture}(150, 50)
\put(-130,-500){\epsfxsize=15cm \epsfbox{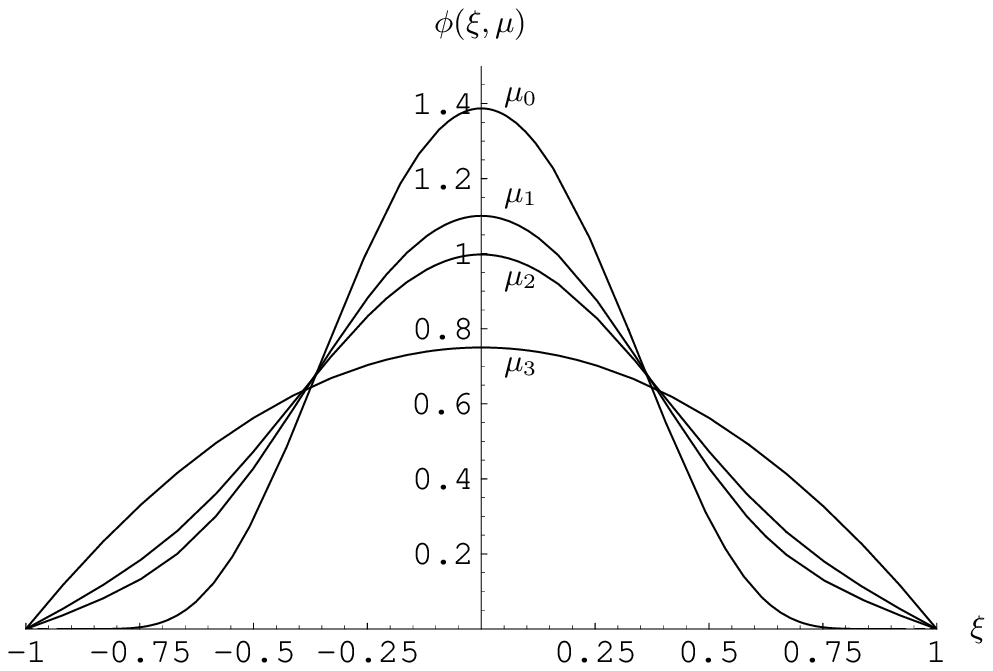}}
\put(-100, -100){{\bf Fig.1:} The LCWF $\phi_L$ (\ref{model}) at scales $\mu_0 =1.2~\mbox{GeV} , \mu_1 = 10 ~\mbox{GeV}, \mu_2 =100 ~\mbox{GeV}, \mu_3 = \infty$.}
\end{picture}
\end{figure}
\vspace*{3cm}

\section{The model for the LCWFs of $J/\Psi$ meson.}

In paper \cite{Braguta:2006wr} the model of leading twist LCWF of $\eta_c$ meson was proposed:
\beq
\phi(\xi, \mu=\mu_0) = c( \beta ) (1- \xi^2) \mbox{exp} \biggl (  -  \frac {\beta} {1-\xi^2} \biggr ),
\label{modeleta}
\eeq
where $c( \beta )$ is a normalization coefficient, the constant $\beta = 3.8 \pm 0.7$, the scale $\mu_0=1.2$ GeV. 
This function allows one to reproduce the results (\ref{reseta}) with rather good accuracy. 
The moments of this wave function are
\beq
\langle \xi^2 \rangle &=& ~0.070 \pm 0.007,  \\
\langle \xi^4 \rangle &=& ~0.012 \pm 0.002, \nonumber \\
\langle \xi^6 \rangle &=& 0.0030 \pm 0.0009. \nonumber  
\eeq
At central value $\beta=3.8$ the constant $c(\beta) \simeq 62$. 

As it was noted in the previous section the accuracy of the calculation does not allow one to distinguish 
leading twist LCWF of $\eta_c$ meson from LCWFs of $J/\Psi$ meson. For this reason the model (\ref{modeleta}) will be used for 
the LCWFs $\phi_{L} (\xi, \mu=\mu_0),~ \phi_{T} (\xi, \mu=\mu_0)$
\beq
\phi_{L} (\xi, \mu=\mu_0) = \phi_{T} (\xi, \mu=\mu_0) = \phi(\xi, \mu=\mu_0).
\label{model}
\eeq
In this expression the functions $\phi_{L} (\xi, \mu=\mu_0),~ \phi_{T} (\xi, \mu=\mu_0)$ are defined at 
scale $\mu=\mu_0$. It is not difficult to calculate these functions at any scale $\mu > \mu_0$ using conformal expansions (\ref{conf_exp}). 
The LCWFs  $\phi_{L} (\xi, \mu)$ at scales $\mu_0 =1.2~\mbox{GeV} , \mu_1 = 10 ~\mbox{GeV}, \mu_2 =100 ~\mbox{GeV}, \mu_3 = \infty$
are shown in Fig. 1. The moments of LCWFs $\phi_{L} (\xi, \mu)$ at scales $\mu_0 = 1.2~\mbox{GeV}, \mu_1 = 10 ~\mbox{GeV}, \mu_2 =100 ~\mbox{GeV}, \mu_3 = \infty$
are presented in second, third, fourth and fifth columns of Table II. The plot and the moments of 
LCWF $\phi_{T} (\xi, \mu)$ will not be shown since this function practically does not deviate 
from $\phi_{L} (\xi, \mu)$.

As was noted in paper \cite{Braguta:2006wr}  model (\ref{model}) has some interesting properties. 
For instance, let us consider the LCWF $\phi_{L} (\xi, \mu)$ (similar consideration can be done for $\phi_{T} (\xi, \mu)$). 
From conformal expansion (\ref{conf_exp}) one can derive the expressions that determine the evolution of the moments:
\beq
\label{xxi}
\langle \xi_L^2 \rangle_{\mu} &=& \frac 1 5 + a^L_2(\mu) \frac {12} {35}, \\ \nonumber
\langle \xi_L^4 \rangle_{\mu} &=& \frac 3 {35} + a^L_2(\mu) \frac {8} {35} + a^L_4(\mu) \frac {8} {77}, \\ \nonumber
\langle \xi_L^6 \rangle_{\mu} &=& \frac 1 {21} + a^L_2(\mu) \frac {12} {77} + a^L_4(\mu) \frac {120} {1001} + a^L_6(\mu) \frac {64} {2145}.
\eeq
Similar relations can be found for any moment. Further let us consider the expression for the second moment $\langle \xi_L^2 \rangle$.
In this paper the value $\langle \xi_L^2 \rangle$ has been found
with some error at scale $\mu = \mu_0$. This means that the value of the coefficient $a^L_2( \mu = \mu_0)$ 
was found with some error. The coefficient $a_2^L$ decreases as scale increases.
So the error in $a_2^L$ and consequently in $\langle \xi_L^2 \rangle$ decreases as scale increases. At 
infinitely large scale there is no error in $\langle \xi_L^2 \rangle$ at all. The calculations show 
that the error $10 \%$ in $\langle \xi_L^2 \rangle$ at scale $\mu=\mu_0$ decreases to $4 \%$ at scale
$\mu = 10$ GeV. Applying relations (\ref{xxi}) it is not difficult to show that similar improvement of the 
accuracy happens for higher moments. The improvement of the 
accuracy allows one to expect that model (\ref{model}) at larger scales will be rather good even if QCD
radiative corrections to results (\ref{resL}), (\ref{resT}) are large.

From Fig. 1 one sees that LCWF at scale $\mu=\mu_0$ practically vanishes 
in the regions $0.75 < |\xi| < 1$. In this region the motion of quark-antiquark pair is relativistic 
and vanishing of LCWF in this region means that at scale $\mu = \mu_0$ charmonium can be considered 
as a nonrelativistic bound state of quark-antiquark pair with characteristic velocity $v^2 \sim 1/ \beta \sim 0.3$.
Further let us regard the function $\phi_L(\xi, \mu=\mu_0)$ as a conformal expansion (\ref{conf_exp}). 
To get considerable suppression of the LCWF in the region $0.75 < |\xi| < 1$ one should require 
fine tuning of the coefficients of conformal expansion $a_n^L ( \mu=\mu_0 )$. The evolution of the constants $a_n^L$ (especially with large $n$)
near $\mu = \mu_0$ is rather rapid (see formulas (\ref{ren}) and (\ref{an_dim})) and if there is fine tuning of the constants
at scale $\mu=\mu_0$ this fine tuning will be rapidly broken at larger scales. This property is well seen 
in Table II and Fig. 1. From Fig. 1 it is seen that there is relativistic tail in the region $0.75 < |\xi| < 1$
for scales $\mu= 10, 100$ GeV which is absent at scale $\mu=\mu_0$. Evidently this tail cannot be regarded
in the framework of NRQCD. This means that, strictly speaking, at some scale 
charmonium can not be considered as nonrelativistic particle and the application of NRQCD to the production of charmonium at large scales
may lead to large error. Although in the above arguments  
the model of LCWF (\ref{model}) was used it is not difficult to understand that 
the main conclusion is model independent.

According to the velocity scaling rule \cite{Bodwin:1994jh} the moments $\langle \xi_L^n \rangle$ of LCWF 
depend on relative velocity as $\sim v^n$. It is not difficult to show that the moments of LCWF (\ref{model})
satisfy these rules. Now let us consider the expressions that allows one to connect 
the coefficients of conformal expansion $a_n^L$ with the moments $\langle \xi_L^n \rangle$.
These expressions for the moments $\langle \xi_L^2 \rangle, \langle \xi_L^4 \rangle, \langle \xi_L^6 \rangle$
are given by formulas (\ref{xxi}). It causes no difficulties to find similar 
expressions for any moment. From expressions (\ref{xxi}) one sees that 
to get velocity scaling rules: $\langle \xi_L^n \rangle \sim v^n$ at some scale one should require 
fine tuning of the coefficients $a_n^L$ at this scale. But, as was already noted above, 
if there is fine tuning of the coefficients at some scale this fine tuning will be 
broken at larger scales. From this one can conclude that velocity scaling rules 
are broken at large scales. 

Consider the moments of LCWF (\ref{model}) at infinite scale. It is not difficult to find that
\beq
\langle \xi_{L, T}^n \rangle_{\mu=\infty} = \frac 3 {(n+1) (n+3)}.  
\eeq
From last equation one can find that $\langle \xi_{L, T}^n \rangle $ does not scale as $v^n$ as 
velocity scale rules \cite{Bodwin:1994jh} require. Thus scaling rules obtained in paper \cite{Bodwin:1994jh}
are broken for asymptotic function. Actually one does not need to set the scale $\mu$ to infinity to break 
these rules. For any scale $\mu > \mu_0$ there is a number $n_0$ for which the moments $\langle \xi_{L, T}^n \rangle,~ n>n_0$
violate velocity scaling rules. This property is a consequence of the following fact:
beginning from some $n=n_0$ the contribution of the relativistic tail of LCWF, that appears at scales $\mu>\mu_0$, 
to the moments becomes considerable. 

The amplitude $T$ of any hard process with charmonium meson production can be written as a convolution 
of LCWF $\Phi( \xi)$ with hard kernel $H(\xi)$ of the process. If one expands this kernel over $\xi$ and substitute 
this expansion to the amplitude $T$ one gets the results:
\beq
T = \int d \xi H(\xi) \Phi(\xi) = \sum_{n} \frac {H^{(n)} (0)}  {n!} \langle \xi^n \rangle.
\label{s}
\eeq
If one takes the scale $\mu \sim \mu_0$ in formula (\ref{s}), than moments $\langle \xi^n \rangle$ 
scale according to the velocity scaling rules $\sim v^n$ and  one gets usual NRQCD expansion of the amplitude.
However due to the presence of the scale of the hard process $\mu_h \gg \mu_0$ there appear large 
logarithms $\log {\mu_h/\mu_0}$ which spoil NRQCD expansion (\ref{s}). To remove this 
large logarithms one should take $\mu \sim \mu_h$. But at large scales velocity scaling rules 
are broken and application of NRQCD is questionable.  

In papers \cite{Bondar:2004sv, Bodwin:2006dm, Ma:2006hc} it was proposed different models of 
LCWFs of $J/\Psi$ and $\eta_c$ mesons. It is interesting to compare the models proposed in 
these papers with model (\ref{model}). Such comparison was done in paper \cite{Braguta:2006wr} 
and it will not be discussed in this paper. 

\begin{table}
$$\begin{array}{|c|c|c|c|c|}
\hline \langle \xi^n \rangle & \phi(\xi, \mu_0=1.2~\mbox{GeV}) & \phi(\xi, \mu_1=10~\mbox{GeV}) & \phi(\xi, \mu_2=100~\mbox{GeV})
& \phi(\xi, \mu_3=\infty )  \\
  \hline
n=2  &  0.070 & 0.12  & 0.14 &  0.20 \\
\hline
n=4  & 0.012 &  0.040 & 0.052 & 0.086 \\
\hline
n=6  & 0.0032 & 0.019 & 0.026 & 0.048  \\
\hline
\end{array}$$
\caption{ The moments of LCWF (\ref{model}) proposed in this paper at scales $\mu_0 = 1.2~\mbox{GeV}, \mu_1 = 10 ~\mbox{GeV}, \mu_2 =100 ~\mbox{GeV}, \mu_3 = \infty$
are presented in second, third, fourth and fifth columns. }
\end{table}

\section{Conclusion.}

In this paper the moments of leading twist light cone wave functions (LCWF) of $J/\Psi$ meson
have been calculated within three approaches. In the first approach  Buchmuller-Tye and Cornell potential models were applied
to the calculation of the moments of LCWFs. In the second approach the moments of LCWFs were calculated in the 
framework of NRQCD. In the third approach the method QCD sum rules was applied to the calculation of the moments. 
The results obtained within these three approaches are in good agreement with each other for the second 
moment $\langle \xi^2 \rangle$. There is a little disagreement between the predictions for the fourth 
moment $\langle \xi^4 \rangle$. The disagreement between the approaches becomes dramatic for the 
sixth moment $\langle \xi^6 \rangle$. The cause of this disagreement consists in the considerable 
contribution of relativistic motion of quark-antiquark pair inside $J/\Psi$ meson to higher moments
which cannot be regarded reliably in the framework of potential models and NRQCD. The approach based on
QCD sum rules is more reliable, especially for higher moments since it does not consider $J/\Psi$-meson 
as a nonrelativistic object. The main problem of QCD sum rules is that since there is no expressions 
of radiative corrections to sum rules one does not know the size these corrections. 
However, one can expect that QCD radiative corrections will not change the results 
for the moments  $n=2$ and $n=4$ dramatically. As to the sixth moment, the contribution 
the QCD radiative corrections in this case may be important. 

The moments of leading twist LCWFs of $J/\Psi$ meson have been compared with the moments 
of LCWF of $\eta_c$.  It was found no significant difference 
between the moments of leading twist LCWF of $\eta_c$ meson and the moments of LCWFs 
$\phi_{L} (\xi, \mu), \phi_{T} (\xi, \mu)$. The difference is within the error of the 
calculation. For this reason the model of LCWF of $\eta_c$ meson was taken as 
a model for leading twist LCWFs $\phi_{L} (\xi, \mu), \phi_{T} (\xi, \mu)$ of $J/\Psi$
meson. As it was shown in paper \cite{Braguta:2006wr} this model has some interesting properties:

1. Due to the evolution (\ref{conf_exp}) the accuracy of the moments obtained within 
 model (\ref{model}) improves as the scale rises. For instance, if the error  
in determination of the moment $\langle \xi_{L, T}^2 \rangle$ is $10 \%$ at scale $\mu=\mu_0=1.2$ GeV, at scale $\mu = 10$ GeV
the error is $4 \%$. For higher moments the improvement of the accuracy is  even better and 
there is no error at all at infinite scale $\mu = \infty$. The improvement of the 
accuracy allows one to expect that model (\ref{model}) will be rather good even after inclusion 
of the QCD radiative corrections. 

2. At scale $\mu \sim \mu_0$ the LCWFs can be considered as wave functions of nonrelativistic object 
with characteristic width $\sim v^2 \sim 0.3$. Due to the evolution, at larger scales relativistic 
tail appears. This tail cannot be considered in the framework of NRQCD and, strictly speaking, at these scales
$J/\Psi$ meson is not a nonrelativistic object. 

3. It was found that due to the presence of high momentum tail in the LCWFs at scales $\mu> \mu_0$ 
there is violation of velocity scaling rules obtained in paper \cite{Bodwin:1994jh}.
More exactly, for any scale $\mu > \mu_0$ there is a number $n_0$ for which the moments $\langle \xi_{L, T}^n \rangle, n>n_0$
violate NRQCD velocity scaling rules.

Actually, the last two properties are properties of real LCWFs of $J/\Psi$ meson. 

The author thanks A.V. Luchinsky and A.K. Likhoded for useful discussion and help in preparing this paper.
This work is partially supported by Russian Foundation of Basic Research under grant 04-02-17530, Russian Education
Ministry grant RNP-2.2.2.3.6646, CRDF grant MO-011-0, president grant MK-1863.2005.02 and Dynasty foundation.

\end{document}